\documentclass[prl,twocolumn,showpacs,preprintnumbers,amsmath,amssymb,superscriptaddress,amsmath]{revtex4-1}
\makeatletter
\renewcommand*\env@matrix[1][c]{\hskip -\arraycolsep
  \let\@ifnextchar\new@ifnextchar
  \array{*\c@MaxMatrixCols #1}}
\makeatother
\usepackage{amsmath,array}
\usepackage{graphicx,subfigure}  
\usepackage{dcolumn}                 
\usepackage{bm}                          
\usepackage{multirow} 
\def\nuebar{\bar{\nu}_{\lowercase{e}}}

\newcommand{\nnbe}{\stackrel{{\scriptscriptstyle (} - {\scriptscriptstyle )}}{\nu}\!\!\!\! _{\rm e}}
\newcommand{\cnb}{C$\nu$B}
\newcommand{\rnb}{C$\nu_{keV}$B}
\newcommand{\snb}{S$\nu$B}
\newcommand{\snu}{{\rm SNU\,}}

\graphicspath{{./figures/}}

\begin{document}
  
\title{Direct Search for keV Sterile Neutrino Dark Matter with a Stable Dysprosium Target}

\author{T. Lasserre}
\email{Corresponding author: thierry.lasserre@cea.fr}
\affiliation{Commissariat \`a l'Energie Atomique et aux Energies Alternatives,
Centre de Saclay, IRFU, 91191 Gif-sur-Yvette, France}
\affiliation{Astroparticule et Cosmologie APC, 10 rue Alice Domon et 
  L\'eonie Duquet, 75205 Paris cedex 13, France}
\affiliation{Institute for Advanced Study, Technische Universit\"at M\"unchen, James-Franck-Str. 1, 85748 Garching, Germany}
\author{K. Altenmueller}
\affiliation{Commissariat \`a l'Energie Atomique et aux Energies Alternatives,
Centre de Saclay, IRFU, 91191 Gif-sur-Yvette, France}
\affiliation{Institute for Advanced Study, Technische Universit\"at M\"unchen, James-Franck-Str. 1, 85748 Garching, Germany}
\affiliation{Physik-Department and Excellence Cluster Universe, Technische Universit\"at M\"unchen, James-Franck-Str. 1, 85748 Garching}
\author{M. Cribier}
\affiliation{Commissariat \`a l'Energie Atomique et aux Energies Alternatives, Centre de Saclay, IRFU, 91191 Gif-sur-Yvette, France}
\affiliation{Astroparticule et Cosmologie APC, 10 rue Alice Domon et
  L\'eonie Duquet, 75205 Paris cedex 13, France}
\author{A. Merle}
\affiliation{Max-Planck-Institut f\"ur Physik (Werner-Heisenberg-Institut), Foehringer Ring 6, 80805 M\"unchen, Germany}
\author{S. Mertens}
\affiliation{Physik-Department and Excellence Cluster Universe, Technische Universit\"at M\"unchen, James-Franck-Str. 1, 85748 Garching}
\affiliation{Max-Planck-Institut f\"ur Physik (Werner-Heisenberg-Institut), Foehringer Ring 6, 80805 M\"unchen, Germany}
\affiliation{Institut f\"ur Kernphysik, Karlsruher Institut f\"ur Technologie (KIT), D-76021 Karlsruhe, Germany}
\author{M. Vivier}
\affiliation{Commissariat \`a l'Energie Atomique et aux Energies Alternatives,
Centre de Saclay, IRFU, 91191 Gif-sur-Yvette, France}

\date{\today}

\begin{abstract}

We investigate a new method to search for keV-scale sterile neutrinos that could account for Dark Matter. Neutrinos trapped in our galaxy could be captured on stable $^{163}$Dy if their mass is greater than 2.83~keV. Two experimental realizations are studied, an integral counting of $^{163}$Ho atoms in dysprosium-rich ores and a real-time measurement of the emerging electron spectrum in a dysprosium-based detector. The capture rates are compared to the solar neutrino and radioactive backgrounds. An integral counting experiment using several kilograms of $^{163}$Dy could reach a sensitivity for the sterile-to-active mixing angle $\sin^2\theta_{e4}$ of $10^{-5}$ significantly exceeding current laboratory limits. Mixing angles as low as $\sin^2\theta_{e4}  \sim 10^{-7}$ / $\rm m_{^{163}\rm Dy}\rm{(ton)}$ could possibly be explored with a real-time experiment. 
\end{abstract}

\maketitle

Although the existence of dark matter (DM) is strongly supported by cosmological observations~\cite{Ade:2015xua}, its nature is unknown. The Standard Model (SM) of elementary particle physics does not provide any suitable option. Hence other candidates have been proposed in several theories beyond the SM~\cite{Cirelli:2015gux}. The most popular are Weakly Interacting Massive Particles (WIMPs) often identified with neutralinos in supersymmetric extensions of the SM~\cite{Jungman:1995df,Gelmini:2006pw,Belanger:2005kh,Gunion:2005rw}. 
WIMPs in the mass range (100~MeV - 10~TeV) would have non-relativistic velocities at the time of galactic structure formation leading to the creation of a great number of galactic-scale structures.
However when compared to $\rm N$-body simulations of structure formation~\cite{Springel:2005nw} some discrepancies arise at scales smaller than 10~kpc. Too few dwarf satellite galaxies have been observed compared to what CDM models predict, leading to the so-called missing satellite problem~\cite{Klypin:1999uc,Moore:1999nt}. Furthermore, CDM simulations also predict too many halos larger than dwarfs compared to observations. However it is hard to explain why haloes of that size would fail producing visible stars, leading to the too-big-to-fail issue~\cite{BoylanKolchin:2011de}. Finally observations also tend to favor a cored profile of galaxies while simulations predict a cuspy matter distribution~\cite{Dubinski:1991bm}.  Although astrophysical feedback effects~\cite{Maccio':2009dx,GarrisonKimmel:2013aq,Geen:2011fj} or refinement of the simulations including baryons~\cite{Brooks:2012vi,Herpich:2013yga} could perhaps solve those CDM issues, no consensus has yet been reached in the community. So far, no direct DM search experiment has conclusively reported a detection~\cite{Cirelli:2015gux}, providing further motivations for in-depth experimental investigation of alternatives. 

Sterile neutrinos appear in several extensions of the SM as right-handed neutral fermions acting as singlets under the SM gauge group, not interacting but mixing with active neutrinos~\cite{Merle:2013gea}. No gauge symmetry forbids the introduction of a Majorana mass term for the right-handed neutrino. Such mass term could have an arbitrary scale~\cite{Majorana:1937vz}. Hence a sterile neutrino with a keV mass scale could exist and account for a great part of DM~\cite{Adhikari:2016bei}. They are referred to as the Cosmic keV Neutrino Background, \rnb, in the following. The lower bound on their mass, with a value around 1~keV, originates from the Pauli principle limiting the number of fermions contained inside a galaxy (Tremaine-Gunn bound~\cite{Tremaine:1979we,Boyarsky:2008ju}). A stringent bound on their mass and mixing is given by X-ray observations since sterile neutrinos could decay with the emission of mono-energetic photons. 
Recently two independent groups reported evidence for a 3.5~keV emission line that could be due to the decay of a 7.1~keV relic neutrino with $\sin^2(2\theta)\sim10^{-10}$~\cite{Boyarsky:2014jta,Bulbul:2014sua}. This observation is being fiercely debated~\cite{Riemer-Sorensen:2014yda,Anderson:2014tza, Jeltema:2014qfa,Iakubovskyi:2015kwa,Malyshev:2014xqa,Urban:2014yda,Tamura:2014mta,Aharonian:2016gzq}.  
Hardly interacting, keV neutrinos could be gradually produced since their corresponding mass eigenstate is partially active and sterile. 
Resonantly enhanced oscillations in the early Universe related to lepton number violation (Shi-Fuller mechanism) could explain cosmological observation and also explain the 7.1~keV sterile neutrino signal~\cite{Shi:1998km, Canetti:2012kh, Abazajian:2014gza}. Other production mechanisms have also been proposed, like singlet scalar decay~\cite{Merle:2013wta, Merle:2015oja,Konig:2016dzg} or Diluton decay~\cite{Patwardhan:2015kga}. In some of these models the DM velocity spectrum significantly deviates from a thermal spectrum and keV sterile neutrinos then could be described neither as solely warm nor cold DM~\cite{Adhikari:2016bei}, thus possibly circumventing Lyman-$\alpha$ constraints~\cite{Merle:2014xpa, Baur:2015jsy}.
Like any non-relativistic particle, keV sterile neutrinos ($\nu_4$) cluster in the gravitational potential wells of galaxies. Assuming that DM neutrinos of mass ${\rm~m_4}$ account for the entire local DM density, $\rho_{\rm DM} \simeq 0.3\pm0.1~{\rm GeV \cdot cm^{-3}}$ \cite{Bovy:2012tw}, their local number density is $\rm n_{\nu_4} \simeq  (300  \pm100)\cdot \,10^3  / {\rm~m_4} (\rm{keV}) \, {\rm cm}^{-3}$,  hence up to a factor 1000 larger than the Cosmic (active) Neutrino Background (\cnb) and roughly 10$^6$ times larger than for WIMPs. Considering an isothermal Milky Way halo we assume the \rnb\,velocity to follow a shifted Maxwellian distribution with an average value of $\rm v_{\nu_4}$=220~km/s~\cite{Freese:1987wu}. \\

The generic electron (anti)neutrino induced $\beta$-capture on a radioactive nucleus $\rm R$, $\nnbe + \rm R \rightarrow \rm R' + \rm e^\pm$, was addressed in~\cite{Cocco:2007za,Lazauskas:2007da}. Due to the positive energy balance $\rm Q_\beta=\rm M_a(\rm R) - \rm M_a(\rm R')>0$~\footnote{$\rm M_a$ denotes the masses of neutral atoms, $\rm M_n$ denotes the nuclear masses.} this exothermic reaction is always allowed independently of the value of the incoming neutrino energy $\rm E_\nu$. This process is therefore appealing for the detection of the \cnb\,which are expected to have today a tiny kinetic energy $\rm T_{C \nu B}\sim 0.5$~meV~\cite{Cocco:2007za,Lazauskas:2007da,2014EPJWC..7100044F}. 
Considering the case of tritium, a relevant \cnb\,experiment necessitate $\sim100$~g of radioactive target material~\cite{Blennow:2008fh,Li:2010sn,Kaboth:2010kf,Faessler:2013jla,Betts:2013uya,Long:2014zva}. This is a huge technical challenge~\cite{Betts:2013uya} when compared to the already large amount of tritium, $\sim50$~$\mu$g, to be used in the KATRIN experiment~\cite{Angrik:2005ep}. 
Nevertheless it is interesting to look at the characteristic signal associated with this process as the emerging electron would create a mono-energetic peak at $\rm T_e = \rm E_0 + \rm m_\nu$, where $\rm T_e$ is the electron kinetic energy, $\rm E_0$ is the endpoint energy of the $\beta$-decay, and $\rm m_{\nu}$ is the effective electron neutrino mass. In the case of the \cnb, a sub-eV energy resolution is required to distinguish this signal from the tail distribution of the $\beta$-decay spectrum. 

Likewise the detection of the \rnb\, through DM $\nu$-capture on $\beta$-decaying nuclei was also considered~\cite{Shaposhnikov:2007cc,Li:2010vy,Long:2014zva}. The \rnb\, induced mono-energetic electron signal would appear at $\rm T_e = \rm E_0 + \rm m_4$, thus comfortly distinguishable with standard nuclear physics technology. However the expected capture rate suffers from a potentially strong suppression factor $\sin^2\theta_{e4}$ induced by the mixing between the sterile and active neutrino components. A mass of radioactive material, on the scale of 100~g for tritium~\cite{Long:2014zva}, 10~kg for $^{106}$Ru~\cite{Li:2010vy}, and 600~tons for $^{163}$Ho~\cite{Li:2011mw}, would thus be required to probe a mixing angle of the order of $\sin^2\theta_{e4} \sim10^{-6}$. While procuring this amount of radioactive material is extremely challenging such a mass is in contrast conceivable with a non-radioactive target material.\\

We now generically consider the capture of keV sterile neutrinos on a stable nucleus $S$ leading to the radioactive daughter nucleus $D$, produced as a positive ion:
\begin{equation}
\nu \, + \, _{Z}\rm S \rightarrow \rm e^- \, + \, _{(Z+1)} \rm D^+ \, .\label{e:uncap} 
\end{equation} 
To be energetically possible without minimum kinetic energy from the incident neutrino, the mass of this hypothetical sterile neutrino must satisfy
\begin{equation}
\rm m_4 \geq \Delta(\rm S) - \Delta(\rm D) - \rm E_b(\rm D^+) + \rm E_b(\rm D) \simeq \rm Q_{\rm EC}^{\rm tab} \, , \label{e:uncap} 
\end{equation} 
where $\rm E_b$ define the binding energies of the orbital electrons. The difference in the total binding energy of the neutral atom and the single positive ion is  $\sim$100~eV and can be neglected. $\Delta$ are the usual tabulated mass excess and $\rm Q_{\rm EC}^{\rm tab}$ denotes the tabulated Q-value assuming $\rm m_{\nu}=0$~\cite{Wang:2012aa}. Thus the \rnb\,capture on stable nuclei can be stimulated by the neutrino mass energy for $-\rm m_{\nu_{4}}<\rm Q_{\beta}\leq 0$. 
\begin{center}
\begin{table}[htb!]
\begin{tabular}{c|c|c|c|c}
\hline\hline
 \multirow{2}{*}{S-R}&  \multirow{2}{*}{$\pi^{\rm P}$} & $\rm Q_{\beta}^{tab}$ & $\mathcal{A}$(\rm S)   & Decay  Mode   \\ 
       &                           &  (keV)          & (\%)    & T$_{1/2}$ (y)  \\ 
 \hline\hline
$^{205}$Tl-Pb & $1/2^{+}\rightarrow5/2^{-}$  & -50.6 & 70.5 & EC ($1.5\cdot 10^{7}$)\\ 
$^{163}$Dy-Ho  & $5/2^{-}\rightarrow7/2^{-}$  & -2.83 & 24.9 & EC (4570) \\ 
\hline 
\end{tabular}
\caption{Stable (S) nuclei with radioactive daughters (R) such that $-50<Q_{\beta^{-}}\leq$0~keV. Spin ($\pi$), parity (P), and  $\rm Q_{\beta}^{tab}$ are given for ground state to ground state transitions. $\mathcal{A}$(S) is the natural abundance of the stable isotope and T$_{1/2}$ is the half-life of the daughter nucleus.}
\label{t:nuc50keV}
\end{table}
\end{center}
Two stable isotopes have a negative $\rm Q_{\beta}^{tab}\gtrsim-50$~ keV (see table~\ref{t:nuc50keV}), $^{205}$Tl and $^{163}$Dy~\cite{Bennett_1984}, the former being studied for assessing the integrated solar neutrino flux during the past few million years~\cite{freedman1976solar,Pavicevic:2012jma}.
Let us  focus on the $\nu$-capture from the nuclear ground state of~$^{163}$Dy to the ground state of~$^{163}$Ho 
\begin{equation}
\begin{split}
 ^{163}{\rm Dy} ({\rm gs}, {\rm I}^\pi=5/2^{-}) + \nu_4 ({\rm~m_4}>2.83 {\rm~keV}) \\
 \rightarrow ^{163}{\rm Ho} ({\rm gs}, {\rm I'}^{\pi '}=7/2^{-}) + e^- ,\label{e:DyHo} 
\end{split} 
\end{equation} 
stimulated by the neutrino mass energy if ${\rm~m_4} >$~2.83~keV~\cite{Eliseev:2015pda}. Subsequently the $^{163}{\rm Ho}$ nuclei decays through electron capture (EC) with a half-life of 4570~years. The production rate of $^{163}$Ho is given by
\begin{equation}
\rm R_{^{163}{\rm Ho}}= N_{^{163}{\rm Dy}} \cdot <\sigma_{c} \rm v_{\nu_4}> \cdot \rm n_{\nu_{4}} \cdot \sin^2\theta_{e4} \, ,
\label{e:DyHoCaptRate} 
\end{equation} 
where $N_{^{163}{\rm Dy}}$ is the number of target~$^{163}$Dy atoms and $<\sigma_{c} v_{\nu_4}>$ denotes the averaged product of the capture cross section and the \rnb\,velocity. 
Following \cite{Lazauskas:2007da} the cross sections of the process~(\ref{e:DyHo}) can be written as
\begin{equation}
\begin{split}
\sigma = \rm G_{\rm F}^2 \cdot \cos^2 \theta_{\rm C} \cdot  {\rm m_e}^{2} / \pi \cdot \left|\mathcal{M}_{nucl}\right|^{2} \cdot  \tfrac{2 \rm I^{'}+1}{2 \rm I+1} \cdot \\
< \tfrac{\rm c}{\rm v_{\nu_4}}\cdot \rm E_{e}\cdot \rm p_{e}\cdot \rm F(\rm E_{e}) >  \, {\rm cm}^{2} ,
\label{e:crosssec}
\end{split} 
\end{equation}
where $\rm G_{\rm F}$ is the Fermi constant, $\theta_{\rm C}$ is the Cabbibo angle, $\rm E_{e}$ and $\rm p_{e}$ are the electron energy and momentum, evaluated with $\rm m_{e}$ as the unit of energy, $\rm F(\rm E_{e})$ is the Fermi function, and $\left|\mathcal{M}_{nucl}\right|$ the nuclear matrix element for reaction~(\ref{e:DyHo}). 
It can be determined by studying the decay of completely ionized $^{163}$Dy$^{66+}$ through bound state $\beta$-decay ($\beta_b$-decay) together with the emission of a monochromatic $\nuebar$~\cite{Daudel:1947}. 
Since the nuclear parts of $\nu$-capture and of $\beta_b$-decay are identical, a measurement of the $\beta_b$-decay probability of bare $^{163}$Dy provides the unknown nuclear matrix element for the transition to the ground state (gs) of $^{163}$Ho~\cite{Bahcall:1961zz}.
Indeed the $\beta_b$-decay was observed by storing bare~$^{163}$Dy$^{66+}$ ions in a heavy-ion storage ring. A half-life of $47\pm4$~days was derived~\cite{Jung:1992pw}, in agreement with the predicted  half-life of 50 days. This leads to $ft \simeq 6300/\left|\mathcal{M}_{nucl}\right|^{2}\sim 10^5$~\cite{Takahashi:1987zz}, a value 6 times higher than the $\rm ft$ value for the $^{163}$Ho EC-decay.
\begin{figure}[ht!]
\begin{center}
\includegraphics[scale=0.46]{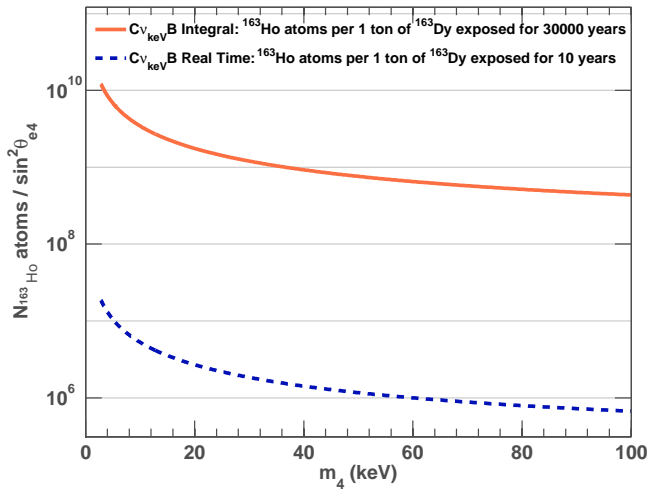}
\caption{\label{f:nho163pertonvskevmass} 
Number of $^{163}$Ho atoms per $\sin^2\theta_{e4}$ produced in 1~tons of $^{163}$Dy through \rnb\, capture on $^{163}$Dy as a function of sterile neutrino mass, ${\rm~m_4}$. Two cases are displayed: an exposure of 10~years for the real-time detection and an exposure of 30000~years for the integral measurement. }
\end{center}
\end{figure}

In the non-relativistic approximation the cross section~(eq. \ref{e:crosssec}) can be expressed as
\begin{equation}
\sigma({\rm m_4}) \simeq 4.8 \cdot 10^{-43} \cdot \rm E_{e}({\rm m_4})\cdot \rm p_{e}({\rm m_4})\cdot \rm F({\rm m_4})   \,{\rm cm}^{2}\, ,
\end{equation}
where the nuclear size and electron screening effects are taken into account in the Fermi function~\cite{shenter:1983}. For $\rm m_4$=5~keV about $10^{7} \times \sin^2\theta_{e4}$ $^{163}$Ho atoms are produced in 1~ton of $^{163}$Dy exposed for 10~years. The number of expected \rnb\,capture as a function of their mass is shown in figure~\ref{f:nho163pertonvskevmass}. The maximal production rate is reached at the threshold, $m_4$=2.83~keV. For higher masses the rate decreases since fewer neutrinos are needed to fill the DM galactic halo.
Taking into account the capture rates given in eq.~(\ref{e:DyHoCaptRate}) the search for \rnb\,appears extremely challenging, mainly because of the huge attenuation factor $\sin^2\theta_{e4} \lesssim 10^{-6}$, as discussed in~\cite{Adhikari:2016bei}. It is therefore appealing to consider an integral experiment where the $^{163}$Dy has been exposed over a geologic time, $\rm t$, enhancing the number of \rnb\,captures given by
\begin{equation}
\begin{split}
N_{^{163}{\rm Ho}}(\rm t,\rm m_4,\sin^2\theta_{e4},{\rm m}_{^{163}{\rm Dy}})= 
\tfrac{ {m}_{^{163}{\rm Dy}} \cdot \mathcal{N}_A}{A_{^{163}{\rm Dy}} \cdot \lambda^{\rm EC}_{^{163}\rm Ho}} \times \\
\sigma(\rm m_4)  <\rm v_{\nu_4}>  \rm n_{\nu_4}  (1-e^{-\lambda^{\rm EC}_{^{163}\rm Ho}  \cdot \rm t}) \cdot \sin^2\theta_{e4} \, ,
\end{split}
\end{equation}
where $A_{^{163}{\rm Dy}}$ is the $^{163}$Dy molar mass and ${m}_{^{163}{\rm Dy}}$ is the target mass.
After an exposure of more than 30000~years an equilibrium is reached between the $^{163}$Ho production and subsequent EC decays. For $m_4$=5~keV about $7 \cdot 10^{9} \cdot \sin^2\theta_{e4}$~atoms of $^{163}$Ho are expected in 1~ton of $^{163}$Dy. This is a gain of three orders of magnitude with respect to a 10~year exposure of a fresh $^{163}$Dy target (see figure~\ref{f:nho163pertonvskevmass}).\\

\begin{figure}[ht!]
\begin{center}
\includegraphics[scale=0.46]{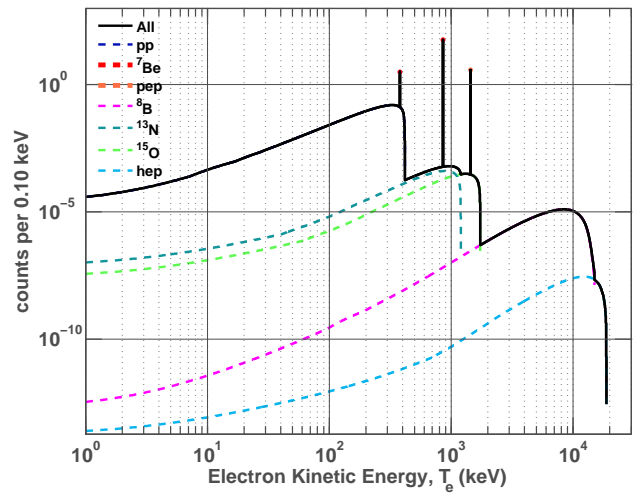}
\caption{\label{f:solnucapDy163spectrum} 
Emitted electron spectra of \snb\, captures in a 1~ton of $^{163}$Dy exposed for 10~years.}
\end{center}
\end{figure}

Solar neutrinos and geoneutrinos with energies above 2.83~keV are also captured on $^{163}$Dy. The geoneutrino $\nu_e$ flux produced by the EC decay of $^{40}$K~\cite{Fiorentini:2002bp} is smaller than the \snb\,, thus we neglect it in the following. However the \snb\,constitutes a background for the search of the \rnb. The capture rates are calculated using the standard solar model including MSW neutrino oscillations~\cite{Bahcall:2004pz}.  A capture rate of 660 \snu is expected, where 1 SNU corresponds to $10^{-36}$ $\nu$-capture per $^{163}$Dy atom per second. It is dominated by the $\rm pp$ (83.5\%) and $^{7}$Be (14.0\%) whereas $\rm pep$, $^{15}$O, $^{19}$F, $^{13}$N and $\rm hep$ solar neutrinos contribute less than 2.5\% (see figure~\ref{f:solnucapDy163spectrum}). 
In a pure $^{163}$Dy target of 1~ton about 400~atoms of $^{163}$Ho are produced during a 10~year exposure. But for $^{163}$Dy exposed for more than 30000~years one expects $\sim 2.7 \cdot 10^5 $  $^{163}$Ho atoms per ton of $^{163}$Dy.

The first proposed detection method consists in counting the number of~$^{163}$Ho atoms in rare-earth ores resulting from the $\nu$-capture on~$^{163}$Dy. 
Besides ore sampling, transport, mineralogical and chemical treatments (based on ion chromatography to separate neighboring lanthanides), the main challenge is the accurate assessment of a small number of $^{163}$Ho atoms out of a large quantity of $^{163}$Dy ore. This could in principle be performed by counting the decays of~$^{163}$Ho as previous radiochemical experiments succeeded to detect approximately one $\nu$-capture per month per ton of target~\cite{Hampel:1998xg}. However, the half-life of $^{163}$Ho, 4570 years, is relatively large and one should then consider other counting methods. $^{163}$Ho atoms counting in a buffer gas loaded magneto-optical trap seems promising~\cite{1367-2630-16-6-063070,Jiang20121}. As for the separation of $^{163}$Ho from other lanthanides, resonance ionization mass spectrometric (RIMS) techniques with lasers could be considered~\cite{kieck:2015}. Accelerator mass spectrometry (AMS) could also be used although it is less sensitive a-priori~\cite{PhysRevLett.93.171103}. Finally neutron activation of $^{163}$Ho and the measurement of the subsequent $\beta$-decay of $^{164}$Ho (29 min~half-life) could also be investigated. In what follows we assume that $^{163}$Ho atoms can be precisely counted.\\

The sensitivity is evaluated by minimizing the following $\chi^2$ function
\begin{equation}
\begin{split}
\chi^2 = -2 \cdot  \left(  \rm N^{obs} - \rm N^{exp} + \rm N^{obs} \cdot \ln{\tfrac{\rm N^{exp}}{\rm N^{obs}}} \right) \\
+ \left(\tfrac{\alpha}{{\sigma_\alpha}}\right)^2 + \left(\tfrac{\beta}{{\sigma_\beta}}\right)^2,
\end{split}
\end{equation}
where $\rm N^{obs}$ and $\rm N^{exp}=(1+\alpha) \cdot (\rm N^{C\nu_{keV}B}+(1+\beta) \cdot \rm N^{S\nu B})$ are the observed and expected number of $^{163}$Ho atoms, respectively. The nuisance parameters ${\alpha,\beta}$ account for both the counting efficiency and the \snb\,uncertainties, ${\sigma_{\alpha}}$ assumed to range from 1 to 20\% and ${\sigma_{\beta}}$ ranging from 1 to 10\%.  After minimizing the $\chi^2$ function over the nuisance parameters ${\alpha,\beta}$, the 90\%~C.L. exclusion contours are computed as a function of $m_4$ and $\sin^2\theta_{e4}$ such that $\Delta \chi^2 = \chi^2(m_4,\sin^2\theta_{e4})-\chi^2_{\rm min}<$ 2.7 since no information on $m_4$ can be retrieved through this integral approach. 
\begin{figure}[ht!]
\begin{center}
\includegraphics[scale=0.46]{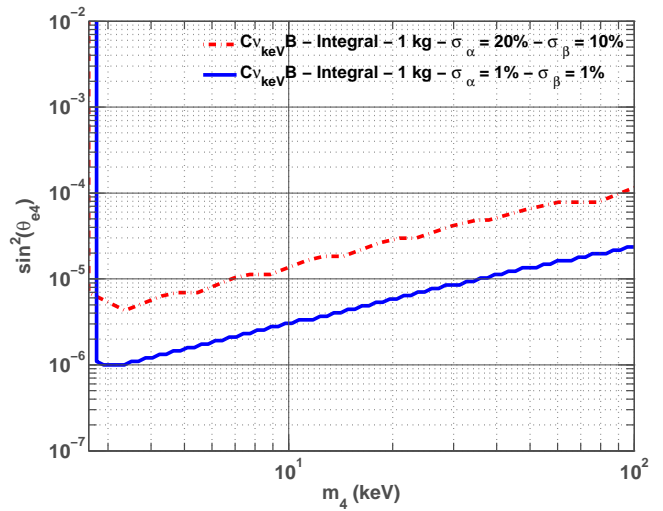}
\caption{\label{f:geosensitivity} 
90\% C.L. sensitivity for \rnb\, integral detection for a target mass of 1~kg, first by assuming an overall counting rate known within 20 to 50 \% and a \snb\,capture rate known with 10\% and second by assuming both the counting rate and the \snb\,capture rate to be known within 1 \%.
}
\end{center}
\end{figure}
Results shown in figure~\ref{f:geosensitivity} indicate that a sensitivity of $\sin^2\theta_{e4} \sim 10^{-5}$ is reachable with a kg-scale target mass. This integral approach is limited by the \snb, however. Indeed, for $\sin^2\theta_{e4} \sim 5 \cdot 10^{-5}$, a similar number of $^{163}$Ho atoms is produced by both the \rnb\,and \snb. Assuming a one percent uncertainty in the knowledge of the \snb\,and the atom counting efficiency the sensitivity could be potentially improved to $\sin^2\theta_{e4} \sim 10^{-6}$. \\

Very likely mineral ore containing dysprosium contains traces of natural uranium and thorium. These radioactive contaminants produce neutrons via spontaneous fission and ($\alpha$,n) reactions on light elements (C, O, Na). The neutron flux induces the capture reaction $^{162}$Er(n,$\gamma$)$^{163}$Er (19~barn) followed by the EC decay of $^{163}$Eu (75~min), leading to $^{163}$Ho. Assuming an thermal neutron flux of $10^{-7}$ n/cm$^2$/s~\cite{Best:2015yma} and taking into account the self-absorption of neutrons on other isotopes, like Gadolinium, we estimate an integral production of less than 1000~$^{163}$Ho atoms for a 1~ton $^{162}$Er target exposed over a geologic time. Assuming a typical dysprosium-rich rock composition, such as the Adamsite~\cite{adamsite}, this production is small compared to the \rnb\,signal for $\sin^2\theta_{e4} > 10^{-6}$. $^{163}$Ho can also be produced by (p,n) or (p,2n) reactions on $^{163}$Dy although these reactions only occur for $>$10~MeV protons. The (p,n) yield is therefore expected to be smaller than for (n,$\gamma$) processes. A detailed estimation of natural backgrounds, beyond the scope of this conceptual study, will require detailed Monte-Carlo simulations using selected Dy-based ore compositions and site-dependent proton and neutron fluxes.\\

\begin{figure}[ht!]
\begin{center}
\includegraphics[scale=0.46]{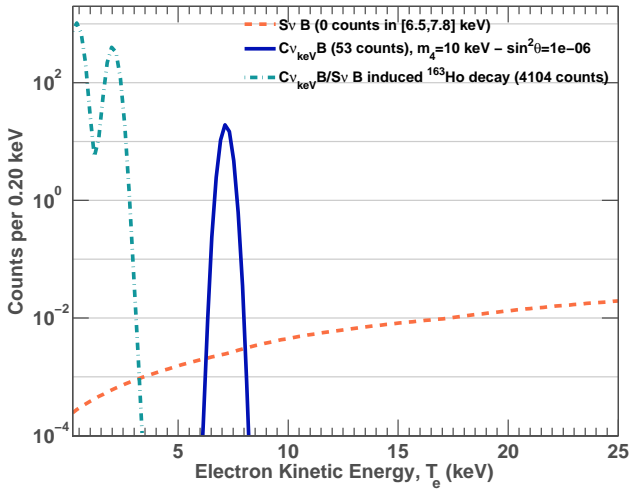}
\caption{\label{f:directexp} 
Electron spectra of \rnb\,and \snb\,captures on $^{163}$Dy and subsequent $^{163}$Ho decays, for an exposure of 100~ton$\cdot$year and a 0.5~keV energy resolution (FWHM).}
\end{center}
\end{figure}

To circumvent limitation due to \snb\,captures in the integral approach another technique, the real-time detection of \rnb\,captures inside an active $^{163}$Dy-based detector, can be considered. The characteristic signal is provided by the mono-energetic electron peak at $\rm T_e = \rm m_4 - 2.83$~keV. Let us consider a detector containing 10~tons of $^{163}$Dy exposed for 10~years (see figure~\ref{f:directexp}). For $\sin^2\theta_{e4} = 10^{-6}$ and $\rm m_4$=10~keV the \rnb\,captures would induce a peak at T$_{\rm e,\rm peak}$=7.2~keV containing 53 electrons, to be discriminated against backgrounds. Compared to the integral case the \snb\,is strongly suppressed since less than 1~\snb\,capture is expected within T$_{\rm e,\rm peak} \pm$~3~$\sigma_T$, Assuming assuming an energy resolution of 0.5~keV (FWHM). 
In this configuration the real-time approach is entangled with the integral approach. Assuming the Dy-based detector was not purified from holmium atoms, the captures \rnb\,and \snb\,integrated over a geologic timescale would produce $^{163}$Ho atoms resulting in subsequent EC-decays affecting the low energy part of the spectrum as displayed in figure~\ref{f:directexp}.

\begin{figure}[ht!]
\begin{center}
\includegraphics[scale=0.46]{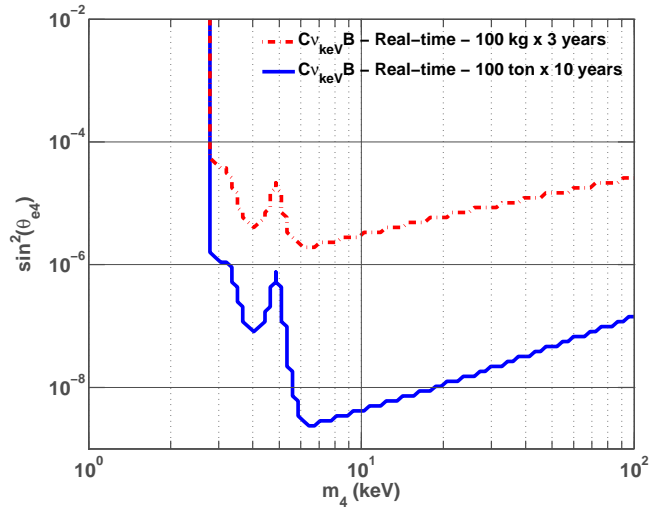}
\caption{\label{f:directsensitivity} 
90\% C.L. sensitivity for \rnb\, real-time detection for two exposures, 300~kg$\cdot$year and 100~ton$\cdot$year (of $^{163}$Dy), assuming an energy resolution of 0.5~keV (FWHM).}
\end{center}
\end{figure}

The sensitivity of the real-time approach is evaluated by minimizing the following $\chi^2$ function
\begin{equation}
\begin{split}
\chi^2 = -2 \cdot  \sum_{i} (  \rm N^{obs}(\rm E_i) - \rm N^{exp}(\rm E_i) \\
 + \rm N^{obs}(\rm E_i) \cdot \ln{\tfrac{\rm N^{\rm exp}(\rm E_i)}{\rm N^{\rm obs}\rm E_i)}}  ),
\end{split}
\end{equation}
where $\rm N^{\rm obs}$ and $\rm N^{\rm exp}= N^{C\nu_{keV}B}+(1+\beta) \cdot N^{S\nu B}+(1+\gamma) \cdot \rm N^{\rm ho}$ are the observed and expected number of counts, respectively. The nuisance parameters ${\beta,\gamma}$ for both the \snb\,capture rate (N$^{S\nu B}$) and the total $^{163}$Ho decays (N$^{\rm ho}$) are left free in this analysis. After minimizing the $\chi^2$ function over the nuisance parameters ${\beta,\gamma}$, the 90\%~C.L. exclusion contours are computed as a function of $\rm m_4$ and $\sin^2\theta_{e4}$ such that $\Delta \chi^2 = \chi^2(m_4,\sin^2\theta_{e4})-\chi^2_{\rm min}<$ 4.6 since both $\rm m_4$ and $\sin^2\theta_{e4}$ could be measured in a real-time experiment. Figure~\ref{f:directsensitivity} shows the expected sensitivity for \rnb\, real-time detection for two different exposures, 300~kg$\cdot$year and 100~ton$\cdot$year, assuming an energy resolution of 0.5~keV (FWHM). Using a few 100~kg of $^{163}$Dy a 90\% sensitivity down to $\sin^2\theta_{e4} \sim10^{-6}$ is attainable for $\rm m_4$ varying from 2.83~to 100~keV. Mixing angles as low as $\sin^2\theta_{e4} \sim10^{-9}$ could in principle be explored with 100~tons of $^{163}$Dy, provided detector backgrounds are less than the \snb\,capture rate. \\

We presented and investigated a new method to search for Cosmic keV Neutrino Background, in the keV mass range, that could account for Dark Matter. This approach aims at the detection of hypothetical keV sterile neutrinos confined inside our galactic halo and is therefore complementary to projects using tritium $\beta$-decay~\cite{Mertens:2014nha,Mertens:2014osa} or xenon~\cite{Campos:2016gjh}. In this new concept the \rnb\,could be captured on stable $^{163}$Dy if sterile neutrino masses are greater than 2.83~keV. Two experimental cases are studied.
First, by an integral experiment in which one counts the number of $^{163}$Ho atoms in a~$^{163}$Dy ore exposed over a geologic time. In this case  the sensitivity to the \rnb\, is strengthened through the accumulation of the captures over the last 30000~years. This approach is limited by the capture of solar neutrinos (\snb), at the level of $\sin^2\theta_{e4} \sim 10^{-6}$.
Second, by a real-time experiment measuring the electron spectrum in a Dy-based detector. The characteristic signal is a mono-energetic peak at $\rm T_e = \rm E_0 + \rm m_4$, thus greatly reducing the impact of the \snb. 

To conclude, experiments using several kilograms of $^{163}$Dy, for the integral approach, and $\sim$100~kg of $^{163}$Dy, for the real-time measurement, could already reach an unprecedented sensitivity of $\sin^2\theta_{e4} \sim 10^{-5}$ in comparison with past laboratory searches. Thanks to the stability of the dysprosium target this experiment is scalable, in principle. Looking into a farther future a cosmological relevant sensitivity of $\sin^2\theta_{e4} \sim 10^{-9}$ could potentially be achievable with 100~tons of $^{163}$Dy exposed for 10~years. Proof-of-concept experiments shall be conducted to experimentally assess backgrounds and technical feasibility, however. \\

Special thanks goes to J.~Rich for many useful inputs and valuable comments. The authors gratefully acknowledge V.~Fischer, L.~Gastaldo, G.~Korschinek, M.~Martini, G.~Mention, W.~Potzel and S.~Sch\"onert for fruitful discussions on $\nu$-capture and detection techniques. We would like to thank B.~Mac~Donough, C.~Jaupart,  M.~Chaussidon, F.~Moynier, and B.~Marty regarding their inputs on geological issues. T.~Lasserre would like to acknowledge the support of the Technische Universit\"at M\"unchen Institute for Advanced Study, funded by the German Excellence Initiative and the European Union Seventh Framework Programme under grant agreement No.~291763, as well as the European Union Marie Curie COFUND program. A.~Merle acknowledges partial support by the Micron Technology Foundation, Inc. A.~Merle furthermore acknowledges partial support by the European Union through the FP7 Marie Curie Actions ITN INVISIBLES (PITN-GA-2011-289442) and by the Horizon 2020 research and innovation programme under the Marie Sklodowska-Curie grant agreements No.~690575 (InvisiblesPlus RISE) and No.~674896 (Elusives ITN).

\bibliography{dy163}

\end{document}